\documentclass{ws-procs961x669}
\usepackage[colorlinks,citecolor=blue,linkcolor=blue,urlcolor=blue]{hyperref}

\begin{document}

\title{Rotation of the crescent image of M87* and polarization of its ESE hotspot}

\author{Krzysztof Nalewajko}

\address{Nicolaus Copernicus Astronomical Center, Polish Academy of Sciences\\
Bartycka 18, 00-716 Warsaw, Poland\\
E-mail: knalew@camk.edu.pl}

\begin{abstract}
The first image of the black hole (BH) M87* obtained by the Event Horizon Telescope (EHT) has the shape of a crescent extending from the E to WSW position angles, with a tentative `ESE hotspot'.
Assuming that the BH spin vector is aligned with both the inner accretion axis and the projected direction of the kpc-scale relativistic jet,
the position of the ESE hotspot is inconsistent with the axisymmetric accretion flow.
Recent polarimetric EHT images of M87* show that the ESE hotspot is essentially unpolarized, which strongly supports its distinct origin.
If the hotspot emission is due to the synchrotron radiation, its depolarization requires either isotropically tangled magnetic fields or an additional Faraday dispersion measure.
The 6-day EHT observing campaign in April 2017 allowed in principle to detect orbital motions advancing by up to $\sim 60^\circ$.
The apparent rotation rate of the major axis of the EHT crescent image is consistent with the rotation rate of the Faraday-corrected polarization angle measured by the ALMA.
However, the counterclockwise (CCW) sense of these rotations is opposite to the clockwise (CW) rotation of the plasma flows implied by the N-S brightness asymmetry, which might indicate that accretion in M87 is retrograde.
\end{abstract}


\bodymatter

\section{Introduction}
\label{sec_intro}

The Event Horizon Telescope (EHT) resolved at $1.3\;{\rm mm}$ the core M87* of the nearby (distance of $D \simeq 16.8\;{\rm Mpc}$) radio galaxy Virgo A (M87) into a crescent image of angular diameter of $\simeq 42\;{\rm \mu as}$, which corresponds to $\simeq 11$ gravitational radii $R_{\rm g} = GM_{\rm BH}/c^2$ for the black hole (BH) mass of $M_{\rm BH} \simeq 6.5\times 10^9 M_\odot$ ($G$ is the gravitational constant, $c$ is the speed of light, $M_\odot$ is the mass of the Sun; $R_{\rm g}/c \simeq 8.9\;{\rm h}$) \citep{2019ApJ...875L...1E,2019ApJ...875L...2E,2019ApJ...875L...3E,2019ApJ...875L...4E,2019ApJ...875L...5E,2019ApJ...875L...6E}.
M87* is thought to be the base of a relativistic jet observed in radio \citep{1999Natur.401..891J,2007ApJ...663L..65C,2018ApJ...855..128W}, optical \citep{1999ApJ...520..621B} and X-ray \citep{2002ApJ...568..133W} bands, propagating in the WNW direction (position angle $\simeq 288^\circ$) beyond the kpc scales.
The inclination angle of the jet axis is estimated at $\simeq 17^\circ$ \citep{2018ApJ...855..128W}.
If the jet is powered by the Blandford-Znajek mechanism \citep{1977MNRAS.179..433B}, its direction of propagation should be aligned with the BH spin.
Accreting plasma in the vicinity of spinning black hole would be forced to swirl with trans-relativistic velocities preferentially along the BH equatorial plane. The resulting difference in the Doppler beaming of emission should result in a roughly N-S brightness asymmetry of the crescent image.
It is in fact observed by the EHT that the S side of the M87* crescent image is significantly brighter than the N side.

The angular extent of the M87* crescent is roughly from the E to WSW position angles.
There is a hint of substructures in the angular brightness profiles that depends significantly on the image reconstruction method.
The observed crescent may consist of the \emph{proper crescent} extending from the SSE to WSW position angles and the \emph{ESE hotspot}.
Using simple axisymmetric models for the geometry of emitting regions, we have previously suggested that the proper crescent can be explained by emitting regions symmetric with respect to the BH spin/jet WNW axis, and that the ESE hotspot (opposite to the WNW jet) would be rather a localized perturbation of the accretion flow \citep{2020A&A...634A..38N}.
Such azimuthal perturbations are commonly observed in the results of GRMHD numerical simulations of magnetized BH accretion flows \citep{2019ApJ...875L...5E,2021MNRAS.502.2023P,2021arXiv210708056S}.

\section{Polarization of the ESE hotspot}

More recently, the EHT Collaboration presented the first resolved polarimetric images of M87* \citep{2021ApJ...910L..12E,2021ApJ...910L..13E}.
These images show linear polarization at the level of $\sim 15\%$.
In our interpretation, this polarization is largely limited the proper crescent, while the ESE hotspot is essentially unpolarized.
Such a striking difference between the polarization properties of these two regions strongly supports their distinct origin.

The very low polarization degree of the ESE hotspot calls into question whether it is produced by the same emission mechanism (presumably synchrotron), whether it is optically thin (or a localized millimeter photosphere), or whether it involves ordered or tangled magnetic fields.
It could also result from depolarization by an additional localized Faraday screen.
Unresolved ALMA polarimetry simultaneous to the EHT campaign was performed in two bands --- $1.3\;{\rm mm}$ and $3\;{\rm mm}$ --- which allowed to estimate the simultaneous rotation measures as $\sim \pm 10^5\;{\rm rad/m^2}$ \citep{2021ApJ...910L..14G}.
Since depolarizing a $1.3\;{\rm mm}$ signal by factor $e$ requires a dispersion measure of $\simeq 4\times 10^5\;{\rm rad/m^2}$, depolarization of the ESE hotspot may require an additional dispersion measure of order $\sim 10^6\;{\rm rad/m^2}$.
Such dispersion measures are plausible in the advection dominated accretion flow (ADAF) models \citep{2006ApJ...640..308M} for the mass accretion rate of $\dot{M} \gtrsim 2\times 10^{-3}M_\odot/{\rm yr}$ \citep{2017MNRAS.468.2214M}.

\section{Rotation of the crescent image and net polarization vector}

The 2017 EHT observing campaign spanned 6 days, resolved images were obtained from data recorded on April 5th, 6th, 10th and 11th.
This time scale can be compared with the orbital period of a source propagating along the innermost stable circular orbit (ISCO) in Schwarzschild\footnote{The effect of BH spin is not included in this basic analysis, but it would be an obvious next level of sophistication.} metric: $P_{\rm ISCO} \simeq 92 R_{\rm g}/c \simeq 34\;{\rm days}$.
Hence, during the 6-day campaign one could expect to observe an orbital advance by $\sim 60^\circ$.
Note that the estimated viewing angle of the M87 jet is sufficiently small that equatorial accretion flows would be well aligned with the plane of the sky.

For comparison, in the case of Sgr~A* ($R_{\rm g}/c \simeq 20\;{\rm s}$; $P_{\rm ISCO} \simeq 31\;{\rm min}$) the VLTI/GRAVITY Collaboration detected orbital motion at the period of $P_{\rm GRAVITY} \simeq 40\;{\rm min} \simeq 1.3P_{\rm ISCO}$ \citep{2018A&A...618L..10G}, which corresponds to a circular Schwarzschild geodesic of radius $R_{\rm circ} \simeq 7.1R_{\rm g}$.

\begin{figure}
\includegraphics[width=\textwidth]{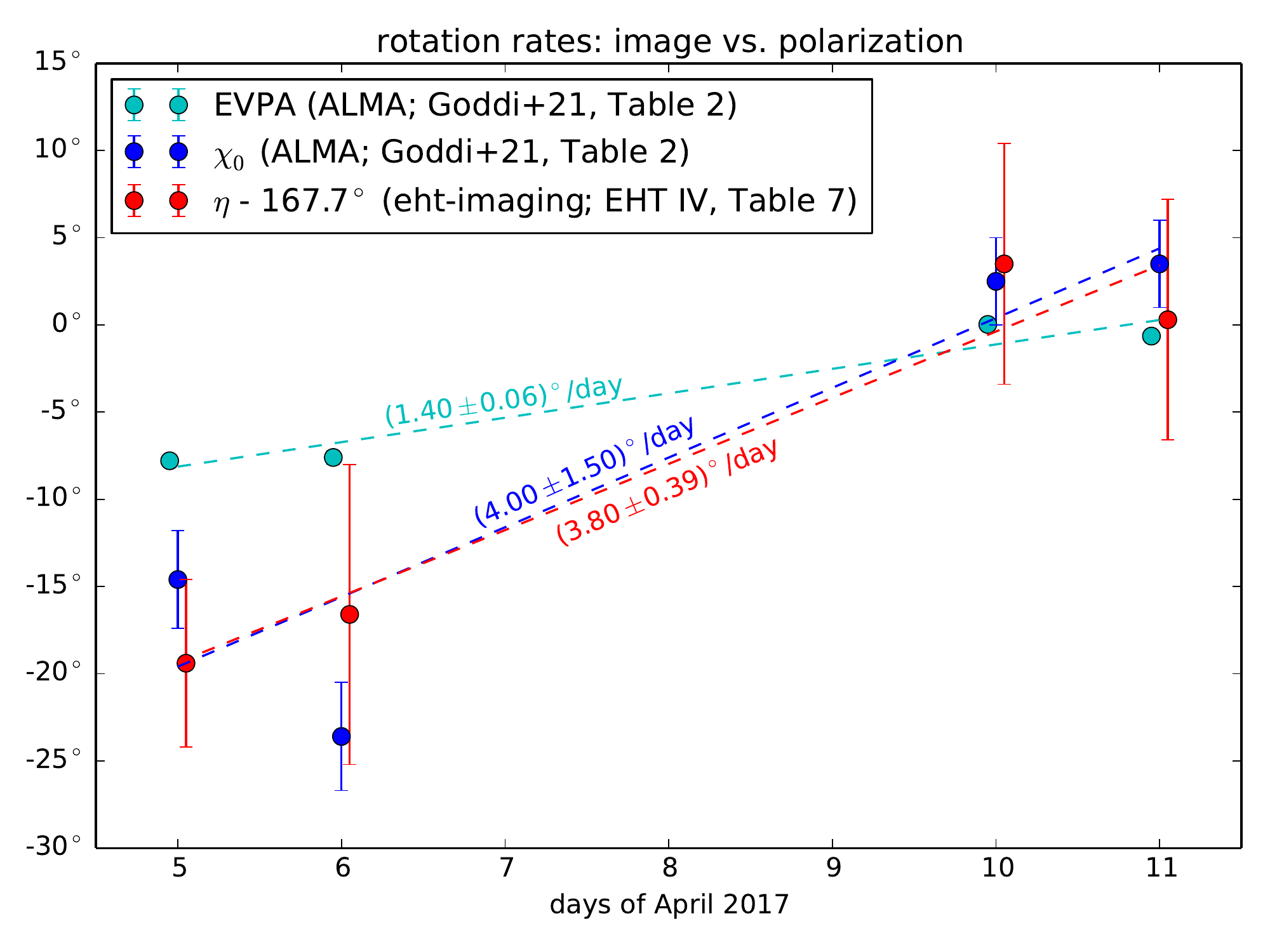}
\caption{Time evolution of the orientation angles of the total-intensity crescent image and net polarization. See the main text for details.}
\label{fig_rotation_rates}
\end{figure}

In Figure \ref{fig_rotation_rates}, we compare two kinds of orientation angles measured independently for each day of the M87 campaign.
First, we consider the mean orientation angles $\eta$ of the crescent image according to the {\tt eht-imaging} algorithm from the Table 7 of EHT Paper IV \citep{2019ApJ...875L...4E}.
This angle increases by $\simeq +23^\circ$ between April 5th and April 10th, and the mean rotation rate is $(3.8 \pm 0.6)^\circ/{\rm day}$.
Second, we consider the net polarization angles from the Table 2 of the ALMA Paper \citep{2021ApJ...910L..14G}: the observed angles EVPA, and the angles $\chi_0$ corrected for Faraday rotation using the simultaneous rotation measure values.
The EVPA increases by $\simeq +7.8^\circ$ between April 5th and April 10th, while $\chi_0$ increases by $\simeq +27^\circ$ between April 6th and April 11th.
The mean rotation rate is $(1.4\pm 0.3)^\circ/{\rm day}$ for the EVPA and $(4.0\pm 1.2)^\circ/{\rm day}$ for the $\chi_0$.
The rotation rate of the Faraday-corrected polarization angle is thus consistent with the rotation rate of the crescent.
The rate of $3.9^\circ/{\rm day}$ corresponds to the orbital period of $P \simeq 92\;{\rm days} \simeq 2.7P_{\rm ISCO}$.
The apparent rotation rate of the crescent image and net polarization would correspond to $R_{\rm circ} \simeq 11.6\,R_{\rm g}$.

The trend of increasing position angle measured from N to E corresponds to the counterclockwise (CCW) rotation on the sky for the relatively outer accretion flow.
On the other hand, since the S side of the M87* image is clearly brighter than the N side, should this be attributed to the trans-relativistic Doppler beaming of the inner accretion flow (funnel wall), in light of the roughly W direction of the projected jet, the inner accretion flow (and the BH itself) should rotate in the clockwise (CW) direction on the sky.
This situation of opposite outer and inner accretion flows corresponds to the bottom left panel of Figure 5 in EHT Paper V \citep{2019ApJ...875L...5E}.
This would suggest that accretion in M87 is retrograde.
And this in turn would constrain the range of GRMHD models listed in Table 2 of EHT Paper V \citep{2019ApJ...875L...5E} to two categories: the high-spin ($a_* = -0.94$) SANE models, and the medium-spin ($a_* = -0.5$) MAD models.

\section*{Acknowledgments}

I thank Brian Punsly for motivating this contribution by inviting the author to speak at the Sixteenth Marcel Grossmann Meeting.
I also thank Agata R{\'o}{\.z}a{\'n}ska and Marek Sikora for discussions.
This work was partially supported by the Polish National Science Centre grant 2015/18/E/ST9/00580.

\bibliography{nalewajko_mg16_v3}


\end{document}